\documentstyle[prd,preprint,tighten,aps,eqsecnum,amssymb,newlfont,epsfig]{revtex}
\begin{document}
\draft
\preprint{
\begin{tabular}{r}
   DFTT 75/98
\\ hep-ex/9901015
\end{tabular}
}
\title{Treatment of the background error in the statistical analysis
of Poisson processes}
\author{C. Giunti}
\address{INFN, Sezione di Torino, and Dipartimento di Fisica Teorica,
Universit\`a di Torino,\\
Via P. Giuria 1, I--10125 Torino, Italy}
\maketitle
\begin{abstract}
The formalism that allows to take into account the
error $\sigma_b$ of the expected mean background $\overline{b}$
in the statistical analysis of
a Poisson process
with the frequentistic method
is presented.
It is shown that the error $\sigma_b$
cannot be neglected if it
is not much smaller than $\sqrt{\overline{b}}$.
The resulting confidence belt is larger that
the one for $\sigma_b=0$,
leading to larger confidence intervals
for the mean $\mu$ of signal events.
\end{abstract}

\pacs{PACS numbers: 06.20.Dk
\\
\\ Published in Phys. Rev. D \textbf{59}, 113009 (1999).
}

\section{Introduction}
\label{Introduction}

The statistical treatment of experimental results
obtained in a Poisson process with background and a small signal
is difficult and controversial.
Two methods are accepted by the Particle Data Group~\cite{PDG98}:
the Bayesian Method and the Unified Approach,
which is a frequentist method
proposed recently by Feldman and Cousins~\cite{Feldman-Cousins-98}
that allows the unified calculation of confidence intervals and upper limits
\emph{with the correct coverage}
(see \cite{Cousins95}).

The Unified Approach
represents a major breakthrough for a satisfactory
statistical treatment of processes with small signals
with the frequentist method.
However,
as already noted by Feldman and Cousins~\cite{Feldman-Cousins-98},
when the number of observed events
in a Poisson process with mean $\mu$ is smaller than the expected background,
the upper limit for $\mu$ obtained with the Unified Approach
decreases rapidly when the background increases.
Hence,
by observing less events than the expected background
an experiment can establish a very stringent upper bound on $\mu$
even if it is not sensitive to such small values of $\mu$.
This problem has been further discussed in Ref.~\cite{Giunti98-poisson},
where an alternative frequentist method has been proposed.
This method yields confidence intervals and upper limits
with all the desirable properties
of those calculated with the Unified Approach
and in addition minimizes the effect
of the observation of less background events than expected.
In the following this method will be called
``Alternative Unified Approach".
The basic features
of the Unified Approach and the Alternative Unified Approach,
which are necessary for the understanding of the present paper,
are reviewed in Section~\ref{Poisson processes with background}.

The original formulation of the Unified Approach~\cite{Feldman-Cousins-98}
and of the
Alternative Unified Approach~\cite{Giunti98-poisson}
for a Poisson process with background
assumed a precise knowledge of the expected mean background.
The aim of this paper is the presentation of
the extension of these approaches
to the case
in which the background is known with a non-negligible error.
This is done in
Sections~\ref{Background with small error}
and \ref{Background with large error},
where the probability to observe a number $n$ of events
in a Poisson process consisting in signal events
with mean $\mu$ and background events with known mean
$b=\overline{b}\pm\sigma_b$
is derived
(in Section~\ref{Background with small error}
we consider the simpler case
$ \sigma_b \lesssim \overline{b}/3 $
and in
Section~\ref{Background with large error}
this constraint is removed),
and
in Section~\ref{Confidence intervals},
where the method for deriving the corresponding confidence intervals
in the Unified Approach and in the Alternative Unified Approach
is presented.
Conclusions are drawn in Section~\ref{Conclusions}.

\section{Poisson processes with background}
\label{Poisson processes with background}

The probability to observe a number $n$
of events in a Poisson process
consisting in signal events with mean $\mu$
and background events with known mean $b$
is
\begin{equation}
P(n|\mu;b)
=
\frac{1}{n!} \ (\mu+b)^n \, e^{-(\mu+b)}
\,.
\label{poisson}
\end{equation}
The classical frequentist method for obtaining the confidence interval
for the unknown parameter $\mu$
is based on Neyman's method to construct a \emph{confidence belt}
\cite{Neyman37}.
This confidence belt is the region in the $n$--$\mu$ plane
lying between the two curves $n_1(\mu;b,\alpha)$ and $n_2(\mu;b,\alpha)$
such that for each value of $\mu$
\begin{equation}
P(n\in[n_1(\mu;b,\alpha),n_2(\mu;b,\alpha)]|\mu;b)
\equiv
\sum_{n=n_1(\mu;b,\alpha)}^{n_2(\mu;b,\alpha)} P(n|\mu;b)
=
\alpha
\,,
\label{CL}
\end{equation}
where
$\alpha$ is the desired confidence level.
The two curves
$n_1(\mu;b,\alpha)$ and $n_2(\mu;b,\alpha)$
are required to be monotonic functions of $\mu$
and can be inverted to yield the corresponding curves
$\mu_1(n;b,\alpha)$ and $\mu_2(n;b,\alpha)$.
Then,
if a number $n_{\mathrm{obs}}$ of events is measured,
the confidence interval for $\mu$ is
$[\mu_2(n_{\mathrm{obs}};b,\alpha),\mu_1(n_{\mathrm{obs}};b,\alpha)]$.
This method guarantees by construction the \emph{correct coverage},
\textit{i.e.}
the fact that the resulting confidence interval
$[\mu_2(n_{\mathrm{obs}};b,\alpha),\mu_1(n_{\mathrm{obs}};b,\alpha)]$
is a member of a set of confidence intervals
obtained with an ensemble of experiments
that
contain the true value of $\mu$ with a probability $\alpha$
(in other words,
$100\alpha\%$
of the confidence intervals in the set contain the true value of $\mu$).

As noted by Cousins
in Ref.~\cite{Cousins95},
Neyman himself pointed out \cite{Neyman37} that
the usefulness of classical confidence intervals lies in the fact that
the experiments in the ensemble do not need to be identical,
but can be real, different experiments.
One can see this fact in a simple way by considering,
for example,
two different experiments that measure the same quantity $\mu$.
The $100\alpha\%$
classical confidence interval obtained from the results of each experiment
belongs to a set of confidence intervals which can be obtained
with an ensemble of identical experiments
and contain the true value of $\mu$ with probability $\alpha$.
It is clear that the sum of these two sets of confidence intervals
is still a set of confidence intervals that contain the true
value of $\mu$ with probability $\alpha$.

In the case of a Poisson process,
since $n$ is an integer,
the relation (\ref{CL})
can only be approximately satisfied and in practice the chosen
\emph{acceptance intervals}
$[n_1(\mu;b,\alpha),n_2(\mu;b,\alpha)]$
are the smallest intervals such that
\begin{equation}
P(n\in[n_1(\mu;b,\alpha),n_2(\mu;b,\alpha)]|\mu;b)
\geq
\alpha
\,.
\label{CLP}
\end{equation}
This choice introduces an overcoverage for some values of $\mu$
and the resulting confidence intervals
are \emph{conservative}.
As emphasized in Ref.~\cite{Feldman-Cousins-98},
conservativeness is an undesirable but unavoidable property
of the confidence intervals in the case of a Poisson process
(it is undesirable because it
implies a loss of power
in restricting the allowed range for the parameter $\mu$).

The construction of Neyman's confidence belt
\emph{is not unique},
because in general there are many different couples of curves
$n_1(\mu;b,\alpha)$ and $n_2(\mu;b,\alpha)$
that satisfy the relation (\ref{CL}).
Hence,
an additional criterion is needed in order to
define uniquely the acceptance intervals
$[n_1(\mu;b,\alpha),n_2(\mu;b,\alpha)]$.
The two common choices are 
\begin{equation}
P(n<n_1(\mu;b,\alpha)|\mu;b)
=
P(n>n_2(\mu;b,\alpha)|\mu;b)
=
\frac{1-\alpha}{2}
\,,
\label{central}
\end{equation}
which leads to
\emph{central confidence intervals}
and
\begin{equation}
P(n<n_1(\mu;b,\alpha)|\mu;b)
=
1-\alpha
\,,
\label{upper}
\end{equation}
which leads to
\emph{upper confidence limits}.
Central confidence intervals are appropriate for the
statistical description of the results of experiments reporting a positive result,
\textit{i.e.} the measurement of a number of events
significantly larger than the expected background.
On the other hand,
upper confidence limits are appropriate for the
statistical description of the results of experiments reporting a negative result,
\textit{i.e.} the measurement of a number of events
compatible with the expected background.
However,
Feldman and Cousins~\cite{Feldman-Cousins-98}
noticed that switching from central confidence level
to upper confidence limits or vice-versa
on the basis of the experimental data
(``flip-flopping'')
leads to undercoverage for some values of $\mu$,
which is a serious flaw for a frequentist method.

Feldman and Cousins \cite{Feldman-Cousins-98}
proposed an ordering principle
for the construction of the acceptance intervals
that is
based on likelihood ratios
and produces an automatic transition
from central confidence intervals to upper limits
when the number of observed events in a Poisson process with background
is of the same order or less than the expected background,
guaranteeing the correct frequentist coverage for all values of $\mu$.
The acceptance interval for each value of $\mu$
is calculated assigning at each value of $n$ a rank
obtained from the relative size of the likelihood ratio
\begin{equation}
R_{\mathrm{UA}}(n|\mu;b)
=
\frac{ P(n|\mu;b) }{ P(n|\mu_{\mathrm{best}};b) }
\,,
\label{RUA}
\end{equation}
where $\mu_{\mathrm{best}}=\mu_{\mathrm{best}}(n;b)$
(for a fixed $b$)
is the non-negative value of $\mu$ that
maximizes the probability
$P(n|\mu;b)$:
\begin{equation}
\mu_{\mathrm{best}}(n;b)
=
\mathrm{max}[0,n-b]
\,.
\label{best}
\end{equation}
For each fixed value of $\mu$,
the rank of each value of $n$
is assigned in order of decreasing value of the ratio $R_{\mathrm{UA}}(n|\mu;b)$:
the value of $n$ which has bigger $R_{\mathrm{UA}}(n|\mu;b)$ has rank one,
the value of $n$ among the remaining ones
which has bigger $R_{\mathrm{UA}}(n|\mu;b)$ has rank two
and so on.
The acceptance interval for each value of $\mu$
is calculated by adding the values of $n$ in increasing order of rank
until the condition (\ref{CLP}) is satisfied.

The automatic transition from two-sided confidence intervals
to upper confidence limits for $ n \lesssim b $
is guaranteed in the Unified Approach by the fact that
$\mu_{\mathrm{best}}$
is always non-negative.
Indeed,
since
$ \mu_{\mathrm{best}}(n{\leq}b;b) = 0 $,
the rank of
$n \leq b$ for $\mu=0$
is one,
implying that the interval
$ 0 \leq n \leq b $
for $\mu=0$
is guaranteed to lie in the confidence belt.

As already noticed by Feldman and Cousins \cite{Feldman-Cousins-98},
when $n \lesssim b$
the upper bound
$\mu_1(n;b,\alpha)$
decreases rather rapidly when $b$ increases
and stabilizes around a value close to 0.8
for large values of $b$.
Hence,
a stringent upper bound for $\mu$
obtained with the Unified Approach
by an experiment that has observed a number of events
significantly smaller than the expected background
is not due to the fact that the experiment is very sensitive to small values of $\mu$,
but to the fact that less background events than expected have been observed.

The Alternative Unified Approach proposed in
Ref.~\cite{Giunti98-poisson} allows
the construction of a classical confidence belt
which has all the desirable features of the one
in the Unified Approach
(\textit{i.e.}
an automatic transition with the correct coverage
from two-sided confidence intervals to
upper confidence limits when the observed number of events
is of the order or less than the expected background)
and in addition minimizes
the decrease of the upper confidence limit $\mu_1(n;b,\alpha)$ for a given $n$
as the mean expected background $b$ increases.
The Alternative Unified Approach is based on
an ordering principle
for the construction of a classical confidence belt
that is implemented as the
Feldman and Cousins ordering principle in the Unified Approach,
but for each value of $\mu$
the rank of each value of $n$
is calculated from the relative size of the likelihood ratio
\begin{equation}
R_{\mathrm{AUA}}(n|\mu;b)
=
\frac{ P(n|\mu;b) }{ P(n|\mu_{\mathrm{ref}};b) }
\,,
\label{RNO}
\end{equation}
where the reference value $\mu_{\mathrm{ref}}=\mu_{\mathrm{ref}}(n;b)$
is taken to be the bayesian expected value for $\mu$:
\begin{equation}
\mu_{\mathrm{ref}}(n;b)
=
\int_0^\infty \mu \, P(\mu|n;b) \, \mathrm{d}\mu
=
n + 1
-
\left( \displaystyle \sum_{k=0}^{n} \frac{k\,b^k}{k!} \right)
\left( \displaystyle \sum_{k=0}^{n} \frac{b^k}{k!} \right)^{-1}
\,.
\label{mu_ref}
\end{equation}
Here $P(\mu|n;b)$
is the bayesian probability distribution for $\mu$
calculated assuming a constant prior
for $\mu\geq0$
(see, for example, \cite{D'Agostini95}):
\begin{equation}
P(\mu|n;b)
=
\frac{1}{n!}
\
\big( \mu + b \big)^n
\
e^{-\mu}
\left( \displaystyle \sum_{k=0}^{n} \frac{b^k}{k!} \right)^{-1}
\,.
\label{bayes}
\end{equation}
The assumption
of a constant prior is arbitrary,
but it seems to be the most natural choice if $\mu$
is the parameter under investigation and there is no prior
knowledge on its value.
Notice also that the arbitrariness induced by the choice of the prior
is of ``second order'' with respect to the dominant arbitrariness
induced by the choice of the method for constructing
the confidence belt.

The obvious inequality
$
\sum_{k=0}^{n} k\,b^k/k!
\leq
n \sum_{k=0}^{n} b^k/k!
$
implies that
$ \mu_{\mathrm{ref}}(n;b) \geq 1 $.
Therefore,
$\mu_{\mathrm{ref}}(n;b)$
represents a reference value for $\mu$ that not only is non-negative,
as desired in order to have an automatic
transition from
two-sided intervals to upper limits,
but is even bigger or equal than one.
This is a desirable characteristic in order to
obtain a weak decrease of the upper confidence limit
for a given $n$
when the expected background $b$ increases.
Indeed,
it has been shown in Ref.~\cite{Giunti98-poisson}
that for
$ n \lesssim b $
the upper bound 
$\mu_1(n;b,\alpha)$
decreases rather weakly when $b$ increases
and stabilizes around a value close to 1.7
for large values of $b$.
This behaviour of $\mu_1(n;b,\alpha)$
is more suitable for the physical interpretation of
experimental results
than the behaviour of $\mu_1(n;b,\alpha)$
in the Unified Approach.
Furthermore,
as shown by the example in Ref.~\cite{Giunti98-poisson},
the upper limits $\mu_1(n;b,\alpha)$ obtained with the Alternative Unified Approach
for $ n \lesssim b $
are are in reasonable agreement with those obtained with the Bayesian Approach.
Hence,
the Alternative Unified Approach
extends the approximate agreement between the Bayesian and frequentist methods
from $n \gg b$ to $ n \lesssim b $
(although the statistical interpretations
of the confidence intervals is different in the two methods).

\section{Background with small error}
\label{Background with small error}

Let us consider an experiment that measures a Poisson process with
an expected background
$ b = \overline{b} \pm \sigma_b $
and a normal probability distribution function for the
mean expected background $b$:
\begin{equation}
f(b;\overline{b},\sigma_b)
=
\frac{ 1 }{ \sqrt{2\pi} \ \sigma_b }
\
\exp\left[ - \frac{ (b-\overline{b})^2 }{ 2 \, \sigma_b^2 } \right]
\,.
\label{normal}
\end{equation}
The importance of $\sigma_b$
can be estimated by comparing it with $\sqrt{\overline{b}}$,
which represents the rms fluctuation of the number of
background events if $ b = \overline{b} $.
If $ \sigma_b \ll \sqrt{\overline{b}} $
the uncertainty of the value of the background is much
smaller than the typical
fluctuation of the number of observed events induced by the background
and can be safely neglected.
Here we consider the possibility that
$ \sigma_b $ is not much smaller than $ \sqrt{\overline{b}} $
and its contribution cannot be neglected.

For simplicity,
in this section
we assume that
$ \sigma_b \lesssim \overline{b}/3 $
and we consider
$b$ varying from $-\infty$ to $+\infty$,
neglecting the small error introduced by considering
negative values of $b$.
This approximation allows a simple analytic solution of all the integrals
involved in the calculation.
The general case with arbitrarily large $\sigma_b$
and $b$ restricted
in the interval $[0,+\infty)$
is treated in Section~\ref{Background with large error}.

If $\mu$ is the mean of true signal events,
the probability $P(n|\mu;\overline{b},\sigma_b)$
to observe $n$ events is given by
\begin{equation}
P(n|\mu;\overline{b},\sigma_b)
=
\int
P(n|\mu;b)
\
f(b;\overline{b},\sigma_b)
\
\mathrm{d}b
\,,
\label{pnm1}
\end{equation}
with the Poisson probability
$P(n|\mu;b)$
given in Eq.(\ref{poisson}).
With the change of variable
$ x = ( b - \overline{b} + \sigma_b^2 ) / \sigma_b $,
the probability
$P(n|\mu;\overline{b},\sigma_b)$
can be written as
\begin{equation}
P(n|\mu;\overline{b},\sigma_b)
=
\frac{1}{n!}
\
\bigg( \mu + \overline{b} - \sigma_b^2 \bigg)^n
\
\exp\left[ - ( \mu + \overline{b} ) + \frac{\sigma_b^2}{2} \right]
\
I_n(\mu,\overline{b},\sigma_b)
\,,
\label{pnm2}
\end{equation}
where
\begin{equation}
I_n(\mu,\overline{b},\sigma_b)
=
\sum_{k=0}^{n}
\left( \begin{array}{c} n \\ k \end{array} \right)
\left(
\frac
{ \sigma_b }
{ \mu + \overline{b} - \sigma_b^2 }
\right)^k
m_k
\,.
\label{In1}
\end{equation}
Here $m_k$
is the $k^{\mathrm{th}}$
central moment of the normal distribution with unit variance,
\begin{equation}
m_k
=
\frac{ 1 }{ \sqrt{2\pi} }
\int_{-\infty}^{+\infty}
x^k
\
e^{ - x^2 / 2 }
\
\mathrm{d}x
\,.
\label{mom1}
\end{equation}
Taking into account that
$
\int x \, e^{ - x^2 / 2 } \, \mathrm{d}x
=
- e^{ - x^2 / 2 }
$,
the integral in Eq.(\ref{mom1})
can be calculated by parts,
yielding
\begin{equation}
m_k
=
\frac{ k! }{ (k/2)! \ 2^{k/2} }
\label{mom2}
\end{equation}
for $k$ even and
$ m_k = 0 $
for $k$ odd.
From Eqs.(\ref{In1}) and (\ref{mom2}),
we obtain
\begin{equation}
I_n(\mu,\overline{b},\sigma_b)
=
\sum_{k=0}^{n/2}
\frac
{ n! }
{ (n-2k)! \ k! \ 2^k }
\
\left(
\frac
{ \sigma_b }
{ \mu + \overline{b} - \sigma_b^2 }
\right)^{2k}
\,.
\label{In2}
\end{equation}

Equation (\ref{pnm2}) gives the formula for the
probability
$P(n|\mu;\overline{b},\sigma_b)$
to observe a number $n$
of events in a Poisson process
consisting in signal events with mean $\mu$
and background events with known mean $b=\overline{b}\pm\sigma_b$,
\textit{i.e.}
it replaces Eq.(\ref{poisson})
if the error $\sigma_b$ of the calculated mean background is not negligible.
The expression (\ref{In2}) for $I_n(\mu,\overline{b},\sigma_b)$
is valid only if
$ \sigma_b \lesssim \overline{b}/3 $,
but, 
as we will see in the next section,
with an appropriate redefinition of $I_n(\mu,\overline{b},\sigma_b)$
the formula (\ref{pnm2}) for $P(n|\mu;\overline{b},\sigma_b)$
is valid for any value of
$\sigma_b$.

\section{Background with large error}
\label{Background with large error}

In this section we present the formalism that allows treatment of cases in which
$\sigma_b$
is arbitrarily large
and $b$ is restricted
to the interval $[0,+\infty)$.
The gaussian probability distribution function
of the mean expected background $b$
normalized in the interval $[0,+\infty)$ is
\begin{equation}
f(b;\overline{b},\sigma_b)
=
\frac{ N }{ \sqrt{2\pi} \ \sigma_b }
\
\exp\left[ - \frac{ (b-\overline{b})^2 }{ 2 \, \sigma_b^2 } \right]
\qquad
( b \geq 0 )
\,,
\label{normalized}
\end{equation}
with the normalization factor $N$ given by\footnote{The
error function is defined by
$
\mathrm{erf}(x)
\equiv
\frac{2}{\sqrt{\pi}}
\,
\int_0^x e^{-x^2} \, \mathrm{d}x
$.}
\begin{equation}
N^{-1}
=
\frac{ 1 }{ 2 }
\left[
1 + \mathrm{erf}\!\left( \frac{ \overline{b} }{ \sqrt{2} \, \sigma_b } \right)
\right]
\,.
\label{N}
\end{equation}
Apart from the error function that must be evaluated numerically,
the integral over $\mathrm{d}b$
in Eq.(\ref{pnm1}) can still be solved analytically.
Indeed,
Eqs.(\ref{pnm2}) and (\ref{In1}) are still valid,
with
\begin{equation}
m_k
=
\frac{ N }{ \sqrt{2\pi} }
\
\int_{x_{\mathrm{min}}}^{+\infty}
x^k \, e^{-x^2/2} \ \mathrm{d}x
\,,
\label{mk0}
\end{equation}
where
\begin{equation}
x_{\mathrm{min}}
=
- \frac{ \overline{b} - \sigma_b^2 }{ \sigma_b }
\,.
\label{xmin}
\end{equation}
The moments (\ref{mk0})
can be calculated by parts, yielding
\begin{equation}
m_{k}
=
\frac{N}{2}
\left[
1
+
\mathrm{erf}\!\left( - \frac{x_{\mathrm{min}}}{\sqrt{2}} \right)
\right]
\frac{ k! }{ (k/2)! \ 2^{k/2} }
+
\frac{ N }{ \sqrt{2\pi} }
\
e^{-x_{\mathrm{min}}^2/2}
\
\frac{ k! }{ (k/2)! }
\left[
\sum_{\ell=0}^{(k/2)-1}
\frac{ \left(\frac{k}{2}-\ell\right)! }{ (k-2\ell)! }
\
\frac{ x_{\mathrm{min}}^{k-2\ell-1} }{ 2^\ell }
\right]
\label{mk1}
\end{equation}
for $k$ even and
\begin{equation}
m_{k}
=
\frac{ N }{ \sqrt{2\pi} }
\
e^{-x_{\mathrm{min}}^2/2}
\
\left(\frac{k-1}{2}\right)!
\left[
\sum_{\ell=0}^{(k-1)/2}
\frac{ 2^\ell }{ (\frac{k-1}{2}-\ell)! }
\,
x_{\mathrm{min}}^{k-2\ell-1}
\right]
\,,
\label{mk2}
\end{equation}
for $k$ odd.
Therefore,
the probability
$P(n|\mu;\overline{b},\sigma_b)$
to observe a number $n$ of events
is given by the formula in Eq.(\ref{pnm2}) with
\begin{eqnarray}
&&
I_n(\mu,\overline{b},\sigma_b)
=
\frac{N}{2}
\left[
1
+
\mathrm{erf}\!\left( \frac{ \overline{b} - \sigma_b^2 }{ \sqrt{2} \, \sigma_b } \right)
\right]
\left[
\sum_{k=0}^{n/2}
\frac
{ n! }
{ (n-2k)! \ k! \ 2^k }
\left(
\frac
{ \sigma_b }
{ \mu + \overline{b} - \sigma_b^2 }
\right)^{2k}
\right]
\nonumber
\\
&&
\hspace{1cm}
+
\frac{ N }{ \sqrt{2\pi} }
\
\exp\left[ - \frac{ ( \overline{b} - \sigma_b^2 )^2 }{ 2 \ \sigma_b^2 } \right]
\nonumber
\\
&&
\hspace{2cm}
\times
\left\{
\sum_{k=0}^{(n-1)/2}
\frac{ n! \ k! }{ (n-2k-1)! \ (2k+1)! }
\left(
\frac
{ \overline{b} - \sigma_b^2 }
{ \mu + \overline{b} - \sigma_b^2 }
\right)^{2k+1}
\sum_{\ell=0}^{k}
\frac{ 2^\ell }{ (k-\ell)! }
\left(
\frac
{ \sigma_b }
{ \overline{b} - \sigma_b^2 }
\right)^{2\ell+1}
\right.
\nonumber
\\
&&
\hspace{2.5cm}
\left.
-
\sum_{k=0}^{n/2}
\frac{ n! }{ (n-2k)! \ k! }
\left(
\frac
{ \overline{b} - \sigma_b^2 }
{ \mu + \overline{b} - \sigma_b^2 }
\right)^{2k}
\sum_{\ell=0}^{k-1}
\frac{ (k-\ell)! }{ \big(2(k-\ell)\big)! \ 2^\ell }
\left(
\frac
{ \sigma_b }
{ \overline{b} - \sigma_b^2 }
\right)^{2\ell+1}
\right\}
\,.
\label{In3}
\end{eqnarray}
These quantities have a cumbersome expression,
but their numerical evaluation with a computer is
not much more difficult than
that of the corresponding quantities in Eq.(\ref{In2})
(however, the calculation of
$I_n(\mu,\overline{b},\sigma_b)$
is rather difficult if
$ \sigma_b^2 > \overline{b} $
because the addenda in Eq.(\ref{In3})
have alternating signs and the roundoff errors
introduced by subtracting large numbers become crucial).

\section{Confidence intervals}
\label{Confidence intervals}

The construction of the confidence belt for the probability (\ref{pnm2})
follows the same procedure described in
Section~\ref{Poisson processes with background}
but now the confidence interval for $\mu$
corresponding to
a number $n_{\mathrm{obs}}$ of observed events is
$
[
\mu_2(n_{\mathrm{obs}};\overline{b},\sigma_b,\alpha)
,
\mu_1(n_{\mathrm{obs}};\overline{b},\sigma_b,\alpha)
]
$,
\textit{i.e.}
it depends on $\overline{b}$ and $\sigma_b$.
The acceptance intervals can be constructed following the same
principles discussed in
Section~\ref{Poisson processes with background}
and
one can construct the confidence belt for central confidence intervals
or upper confidence limits,
or the confidence belt in the Unified Approach
or in the Alternative Unified Approach.
This section is devoted to the presentation of the formalism
for the implementation of 
the Unified Approach
and
of the Alternative Unified Approach.
As an example, we will consider
$\overline{b}=3$
and $\sigma_b=0,1,1.8$.

The quantity $\mu_{\mathrm{best}}(n;\overline{b},\sigma_b)$
in the Unified Approach
is the value of $\mu$ that maximizes $P(n|\mu;\overline{b},\sigma_b)$
and
the acceptance interval for each value of $\mu$
is calculated assigning at each value of $n$ a rank
obtained from the relative size of the ratio
\begin{equation}
R_{\mathrm{UA}}(n|\mu;\overline{b},\sigma_b)
=
\frac
{ P(n|\mu;\overline{b},\sigma_b) }
{ P(n|\mu_{\mathrm{best}};\overline{b},\sigma_b) }
\,.
\label{RUA1}
\end{equation}
The value of
$\mu_{\mathrm{best}}(n;\overline{b},\sigma_b)$
can be easily calculated by hand
for $n=0,1,2$,
whereas
for higher values of $n$
it can be calculated numerically.
The resulting 90\% CL
confidence belts for
$\overline{b}=3$
and
$\sigma_b=0,1,1.8$
are plotted in Fig.~\ref{fig1}.
We have checked that the confidence belt
for $ \sigma_b \lesssim 0.2 $
practically coincides with the one for $\sigma_b=0$,
confirming the prediction that the contribution of
$\sigma_b$ is negligible if
$ \sigma_b \ll \sqrt{\overline{b}} $.

In Fig.~\ref{fig1},
the confidence belt for $\sigma_b=1$
has been obtained with the formulas presented in
Section \ref{Background with small error},
that are valid for $ \sigma_b \lesssim \overline{b}/3 $,
whereas the confidence belt for $\sigma_b=1.8$
has been obtained with the formulas presented in
Section \ref{Background with large error},
which are valid for any value of $\sigma_b$.
We have checked that the confidence belt for $\sigma_b=1$
calculated with the formulas presented in
Section \ref{Background with large error}
practically coincides with the one shown in Fig.~\ref{fig1}.

From Fig.~\ref{fig1}
one can see that
the broadness of the confidence belt increases with $\sigma_b$.
This is due to the fact that the integral in Eq.(\ref{pnm1})
has the effect of flattening the probability
$P(n|\mu;\overline{b},\sigma_b)$
as a function of $n$
for fixed $\mu$
with respect to
$P(n|\mu;\overline{b},\sigma_b=0)$
and
this flattening effect increases with the size of $\sigma_b$.
The shift of the borders of the confidence belt
as $\sigma_b$ increases is not always monotonic
because of the unavoidable overcoverage
caused by the fact that $n$ is an integer
(see Section~\ref{Poisson processes with background}). 

The lower value of $\mu$ for which $n=0$
is out of the confidence belt in Fig.~\ref{fig1}
is lower for $\sigma_b=1.8$
than for
$\sigma_b=0$ and $\sigma_b=1$.
This is caused by the fact that the ratio (\ref{RUA1}) for $n=0$
does not depend on $\sigma_b$.
Indeed, from Eqs.(\ref{pnm2}) and (\ref{In3}) we have
\begin{equation}
P(n=0|\mu;\overline{b},\sigma_b)
=
\frac{N}{2}
\left[
1
+
\mathrm{erf}\!\left( \frac{ \overline{b} - \sigma_b^2 }{ \sqrt{2} \, \sigma_b } \right)
\right]
\exp\left[ - ( \mu + \overline{b} ) + \frac{\sigma_b^2}{2} \right]
\,.
\label{p0}
\end{equation}
Therefore,
$\mu_{\mathrm{best}}(n=0;\overline{b},\sigma_b)=0$
and
\begin{equation}
R_{\mathrm{UA}}(n=0|\mu;\overline{b},\sigma_b)
=
e^{-\mu}
\,.
\label{RUAn0}
\end{equation}
On the other hand,
the ratio
$R_{\mathrm{UA}}(n|\mu;\overline{b},\sigma_b)$
for $n>0$
increases with $\sigma_b$
because of the flattening of
$P(n|\mu;\overline{b},\sigma_b)$
as a function of $n$.
Hence,
the rank of $n=0$ for each value of $\mu$
decreases with the increasing of $\sigma_b$,
causing the peculiar behaviour of the upper bound
$\mu_1(n=0;\overline{b},\sigma_b,\alpha)$
as a function of $\sigma_b$
exemplified in Fig.~\ref{fig1}.
Since the possibility to set a smaller upper bound
on $\mu$ for larger $\sigma_b$
as a consequence of the observation of $n=0$ events
is undesirable from the physical point of view,
we think that in this case the physical interpretation of
the experimental result should be very cautious,
waiting for a better understanding of the background.

In the Alternative Unified Approach
the acceptance interval for each value of $\mu$
is calculated assigning at each value of $n$ a rank
obtained from the relative size of the ratio
\begin{equation}
R_{\mathrm{AUA}}(n|\mu;\overline{b},\sigma_b)
=
\frac
{ P(n|\mu;\overline{b},\sigma_b) }
{ P(n|\mu_{\mathrm{ref}};\overline{b},\sigma_b) }
\,,
\label{RNO1}
\end{equation}
where
the reference value
$\mu_{\mathrm{ref}}=\mu_{\mathrm{ref}}(n;\overline{b},\sigma_b)$
is the bayesian expected value for $\mu$.
In order to calculate analytically the value of $\mu_{\mathrm{ref}}(n;\overline{b},\sigma_b)$,
it is convenient to write the probability (\ref{pnm2}) as
\begin{equation}
P(n|\mu;\overline{b},\sigma_b)
=
\exp\left[ - ( \mu + \overline{b} ) + \frac{\sigma_b^2}{2} \right]
\
\sum_{k=0}^{n}
\frac{ \mu^{n-k} }{ (n-k)! }
\
J_k(\overline{b},\sigma_b)
\,,
\label{pnm3}
\end{equation}
with
\begin{equation}
J_k(\overline{b},\sigma_b)
\simeq
\sum_{j=0}^{k/2}
\frac
{ ( \overline{b} - \sigma_b^2 )^{k-2j} \, \sigma_b^{2j} }
{ (k-2j)! \ j! \ 2^j }
\label{jeik1}
\end{equation}
for
$ \sigma_b \lesssim \overline{b}/3 $
and
\begin{eqnarray}
&&
J_k(\overline{b},\sigma_b)
=
\frac{N}{2}
\left[
1
+
\mathrm{erf}\!\left( \frac{ \overline{b} - \sigma_b^2 }{ \sqrt{2} \, \sigma_b } \right)
\right]
\left(
\sum_{j=0}^{k/2}
\frac
{ ( \overline{b} - \sigma_b^2 )^{k-2j} \ \sigma_b^{2j} }
{ (k-2j)! \ j! \ 2^j }
\right)
\nonumber
\\
&&
\hspace{1cm}
+
\frac{ N }{ \sqrt{2\pi} }
\
\exp\left[ - \frac{ ( \overline{b} - \sigma_b^2 )^2 }{ 2 \ \sigma_b^2 } \right]
\nonumber
\\
&&
\hspace{2cm}
\times
\left\{
\sum_{j=0}^{(k-1)/2}
\frac{ j! }{ (2j+1)! \ (k-2j-1)! }
\sum_{\ell=0}^{j}
\frac{ 2^\ell \ ( \overline{b} - \sigma_b^2 )^{k-2\ell-1} \ \sigma_b^{2\ell+1} }{ (j-\ell)! }
\right.
\nonumber
\\
&&
\hspace{2.5cm}
\left.
-
\sum_{j=0}^{k/2}
\frac{ 1 }{ (k-2j)! \ j! }
\sum_{\ell=0}^{j-1}
\frac{ (j-\ell)! \ ( \overline{b} - \sigma_b^2 )^{k-2\ell-1} \ \sigma_b^{2\ell+1} }{ \big(2(j-\ell)\big)! \ 2^\ell }
\right\}
\label{jeik2}
\end{eqnarray}
for arbitrarily large $\sigma_b$.
For the bayesian probability distribution function for $\mu$
with a constant prior,
\begin{equation}
P(\mu|n;\overline{b},\sigma_b)
=
\frac
{ P(n|\mu;\overline{b},\sigma_b) }
{ \int_0^\infty P(n|\mu;\overline{b},\sigma_b) \ \mathrm{d}\mu }
\,,
\label{bayes1}
\end{equation}
one obtains
\begin{equation}
P(\mu|n;\overline{b},\sigma_b)
=
e^{-\mu}
\left(
\sum_{k=0}^{n}
\frac{ \mu^{n-k} }{ (n-k)! }
\
J_k(\overline{b},\sigma_b)
\right)
\left(
\sum_{k=0}^{n}
J_k(\overline{b},\sigma_b)
\right)^{-1}
\,.
\label{bayes2}
\end{equation}
Hence,
the reference value $\mu_{\mathrm{ref}}(n;\overline{b},\sigma_b)$,
which is the bayesian expected value for $\mu$,
is given by
\begin{equation}
\mu_{\mathrm{ref}}(n;\overline{b},\sigma_b)
=
n + 1
-
\left(
\sum_{k=0}^{n}
k
\
J_k(\overline{b},\sigma_b)
\right)
\left(
\sum_{k=0}^{n}
J_k(\overline{b},\sigma_b)
\right)^{-1}
\,.
\label{mu_ref1}
\end{equation}
If $ \sigma_b \lesssim \overline{b}/3 $
the quantities
$J_k(\overline{b},\sigma_b)$
are given by Eq.(\ref{jeik1})
and one can see that they are all positive.
Hence,
the inequality
$
\sum_{k=0}^{n}
k
\
J_k(\overline{b},\sigma_b)
\leq
n
\sum_{k=0}^{n}
J_k(\overline{b},\sigma_b)
$
implies that
$\mu_{\mathrm{ref}}(n;\overline{b},\sigma_b)\geq1$
as in the case $\sigma_b=0$
(see Eq.(\ref{mu_ref})
and the following discussion).
On the other hand,
the general formula (\ref{jeik2})
allows
$J_k(\overline{b},\sigma_b)$
to be negative and
$\mu_{\mathrm{ref}}(n;\overline{b},\sigma_b)$
is not guaranteed to be larger than one
if $ \sigma_b \gtrsim \overline{b}/3 $.

The 90\% CL
confidence belts in the Alternative Unified Approach for
$\overline{b}=3$
and
$\sigma_b=0,1,1.8$
are plotted in Fig.~\ref{fig2}.
One can see again that the broadness of the confidence belt
increases with $\sigma_b$.
The behaviour of the upper bound
$\mu_1(n=0;\overline{b},\sigma_b,\alpha)$
as a function of $\sigma_b$
is similar to the one obtained in the Unified Approach
and the same caveats apply
to the physical interpretation of the observation of $n=0$ events.

\section{Conclusions}
\label{Conclusions}

We have presented the formalism that allows the
error $\sigma_b$ of the calculated mean background $\overline{b}$ in the
statistical analysis of a Poisson process with the frequentistic method to
be taken into account.
This error must be taken into account if it is not much smaller
than $\sqrt{\overline{b}}$,
which represents
the rms fluctuation of the number of
background events.

We have considered in particular
the Unified Approach~\cite{Feldman-Cousins-98}
and the Alternative Unified Approach~\cite{Giunti98-poisson},
that guarantee by construction a correct frequentist coverage.
We have shown that the broadness of the classical confidence belt
increases with $\sigma_b$,
leading to an increase of the confidence intervals
for the mean $\mu$ of signal events.
The only exception to this behaviour is represented
by the upper bound
$\mu_1(n=0;\overline{b},\sigma_b,\alpha)$,
which decreases with the increasing of $\sigma_b$
for large values of $\sigma_b$
in both approaches.
Hence,
the physical interpretation of the observation of $n=0$ events
when $\sigma_b$ is large
should be very cautious
and the effort towards a better understanding
of the background should receive high priority.

\acknowledgements

I would like to thank S. Yellin for suggesting the simplest way to perform the integral in
Eq.(\ref{pnm1}).

\begin{figure}[h]
\refstepcounter{figure}
\label{fig1}
Fig.~\ref{fig1}.
90\% CL
confidence belts in the Unified Approach for
$\overline{b}=3$
and
$\sigma_b=0,1,1.8$.
\end{figure}

\begin{figure}[h]
\refstepcounter{figure}
\label{fig2}
Fig.~\ref{fig2}.
90\% CL
confidence belts in the Alternative Unified Approach for
$\overline{b}=3$
and
$\sigma_b=0,1,1.8$.
\end{figure}

\newpage

\begin{minipage}[p]{0.95\linewidth}
\begin{center}
\mbox{\epsfig{file=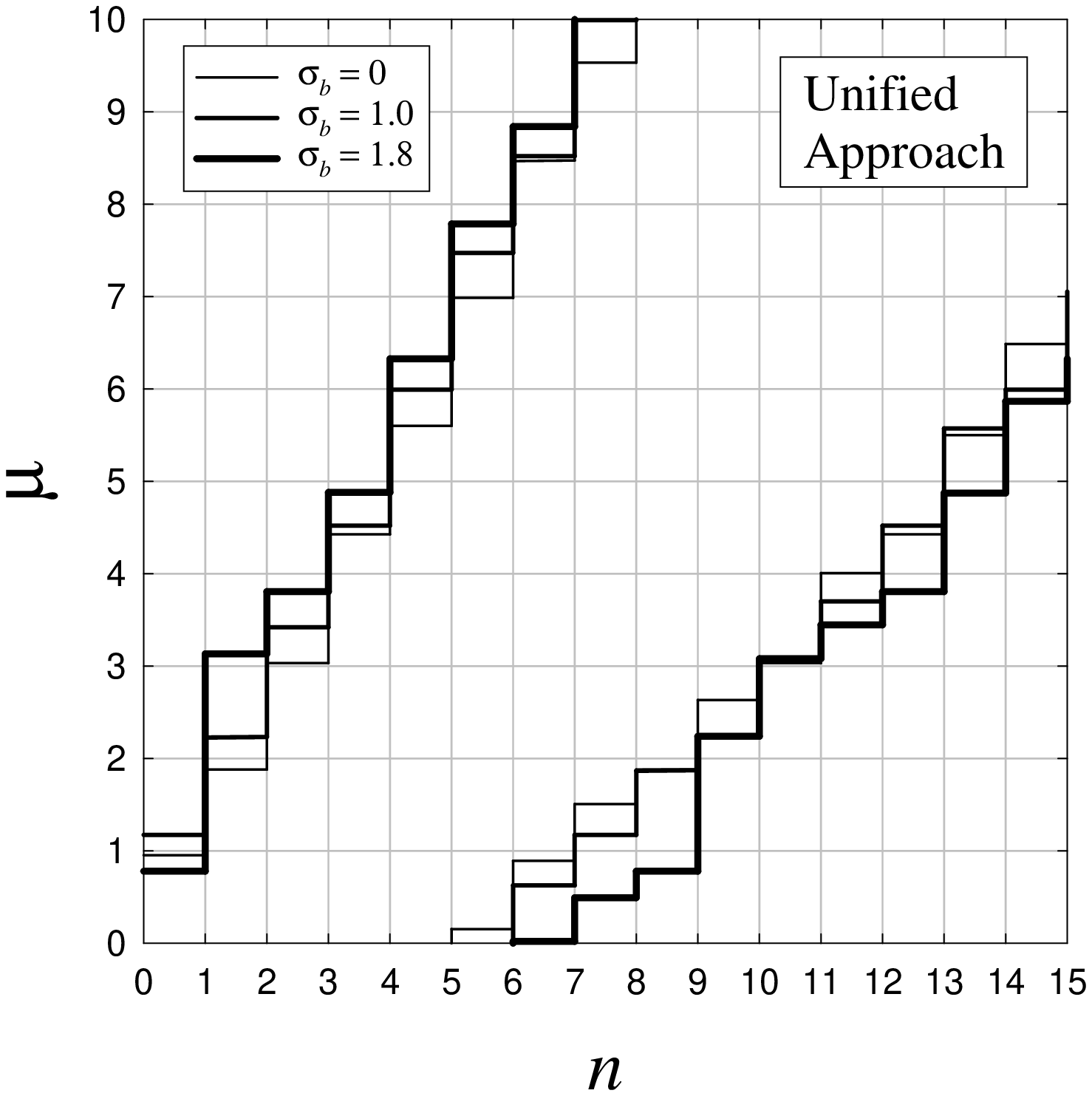,width=0.95\linewidth}}
\end{center}
\end{minipage}
\begin{center}
\Large Figure \ref{fig1}
\end{center}

\begin{minipage}[p]{0.95\linewidth}
\begin{center}
\mbox{\epsfig{file=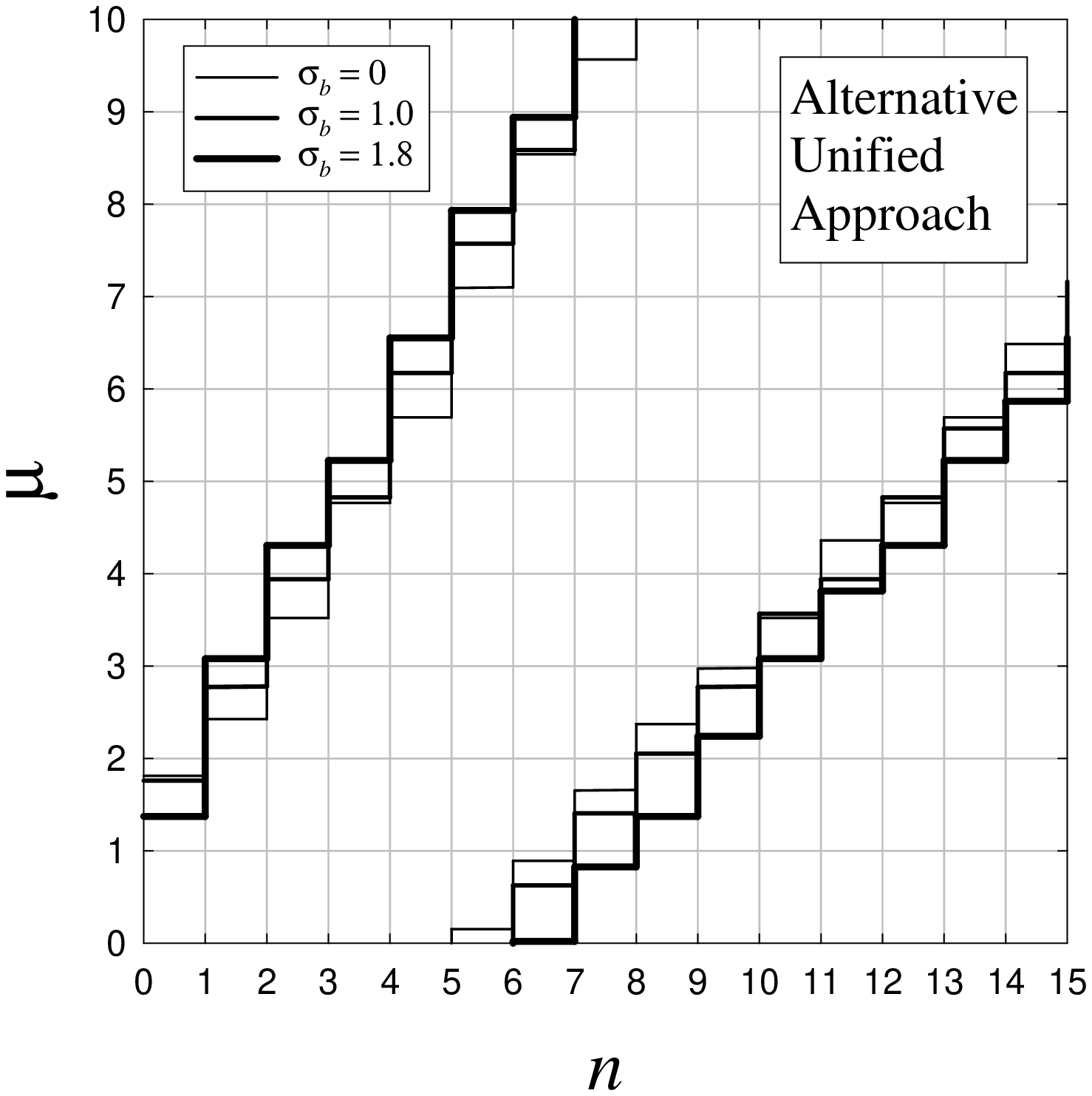,width=0.95\linewidth}}
\end{center}
\end{minipage}
\begin{center}
\Large Figure \ref{fig2}
\end{center}


\begin{references}

\bibitem{PDG98}
C. Caso \textit{et al.},
Eur. Phys. J. C \textbf{3}, 1 (1998).

\bibitem{Feldman-Cousins-98}
G.J. Feldman and R.D. Cousins,
Phys. Rev. D \textbf{57}, 3873 (1998).

\bibitem{Cousins95}
R.D. Cousins,
Am. J. Phys. \textbf{63}, 398 (1995).

\bibitem{Giunti98-poisson}
C. Giunti,
Phys. Rev. D \textbf{59}, 053001 (1999).

\bibitem{Neyman37}
Philos. Trans. R. Soc. London Sect. A \textbf{236}, 333 (1937),
reprinted in
\textit{A selection of Early Statistical Papers on J. Neyman},
University of California, Berkeley, 1967, p.~250.

\bibitem{D'Agostini95}
G. D'Agostini,
preprint DESY 95-242 (hep-ph/9512295).

\end{references}
\end{document}